\documentclass[twocolumn]{aastex7}

\hypersetup{linkcolor=red,citecolor=blue,filecolor=cyan,urlcolor=blue}

\usepackage{graphicx,color}
\usepackage{amssymb}
\usepackage{amsmath}
\usepackage{url}
\usepackage{natbib}
\usepackage{txfonts}
\usepackage{multirow}
\usepackage{array}
\usepackage{rotating}
\usepackage{sidecap}
\usepackage{hyperref}
\usepackage{epstopdf}
\usepackage{footnote}
\usepackage{tabularx}
\usepackage{booktabs}
\usepackage{longtable}
\usepackage{tabu}
\usepackage{longtable}

\newcommand{\FeX}{\ion{Fe}{10}}
\newcommand{\FeIX}{\ion{Fe}{9}}

\newcommand{\kms}{km~s$^{-1}$}
\newcommand{\degree}{\ensuremath{^\circ}}

\newcommand{\sdo}{\textit{SDO}}

\newcommand{\hri}{HRI$_{EUV}$}

\begin{document}

\title  {Buildup, Explosion, and Untwisting of a Solar Active Region Jet Observed with Solar Orbiter, IRIS, and SDO}







\author[0000-0001-7620-362X]{Navdeep K. Panesar}
\affiliation{Lockheed Martin Solar and Astrophysics Laboratory, 3251 Hanover Street, Bldg. 203, Palo Alto, CA 94306, USA}
\affiliation{SETI Institute, 339 Bernardo Ave, Mountain View, CA 94043, USA}
\email{panesar@lmsal.com}

\author[0000-0003-1281-897X]{Alphonse C. Sterling}
\affil{NASA Marshall Space Flight Center, Huntsville, AL 35812, USA}
\email{alphonse.sterling@nasa.gov}
\author[0000-0002-5691-6152]{Ronald L. Moore}
\affil{Center for Space Plasma and Aeronomic Research (CSPAR), UAH, Huntsville, AL 35805, USA}
\affil{NASA Marshall Space Flight Center, Huntsville, AL 35812, USA}
\email{ronald.l.moore@nasa.gov}
\author[0000-0001-7817-2978]{Sanjiv K. Tiwari}
\affil{Lockheed Martin Solar and Astrophysics Laboratory, 3251 Hanover Street, Bldg. 203, Palo Alto, CA 94306, USA}
\affil{Bay Area Environmental Research Institute, NASA Research Park, Moffett Field, CA 94035, USA}
\email{tiwari@lmsal.com}

\author[0000-0003-4052-9462]{David Berghmans}
 \affiliation{Solar-Terrestrial Centre of Excellence—SIDC, Royal Observatory of Belgium, Ringlaan -3- Av. Circulaire, B-1180 Brussels, Belgium}
 \email{david.berghmans@oma.be}
 
 \author[0000-0002-2542-9810]{Andrei Zhukov}
 \affiliation{Solar-Terrestrial Centre of Excellence—SIDC, Royal Observatory of Belgium, Ringlaan -3- Av. Circulaire, B-1180 Brussels, Belgium}
 \affiliation {Skobeltsyn Institute of Nuclear Physics, Moscow State University, 119992 Moscow, Russia}
 \email{Andrei.Zhukov@sidc.be}
 
 \author[0000-0003-4105-7364]{Marilena Mierla}
 \affiliation{Solar-Terrestrial Centre of Excellence—SIDC, Royal Observatory of Belgium, Ringlaan -3- Av. Circulaire, B-1180 Brussels, Belgium}
 \affiliation{Institute of Geodynamics of the Romanian Academy, Bucharest, Romania}
 \email{marilena.mierla@oma.be}
 
 \author[0000-0002-5022-4534]{Cis Verbeeck}
 \affiliation{Solar-Terrestrial Centre of Excellence—SIDC, Royal Observatory of Belgium, Ringlaan -3- Av. Circulaire, B-1180 Brussels, Belgium}
 \email{francis.verbeeck@oma.be}

\author{Koen Stegen}
\affiliation{Solar-Terrestrial Centre of Excellence—SIDC, Royal Observatory of Belgium, Ringlaan -3- Av. Circulaire, B-1180 Brussels, Belgium}
\email{koen.stegen@oma.be}


\begin{abstract}

 We present detailed analysis of an active region coronal jet accompanying a minifilament eruption that is fully captured and well-resolved in high spatial resolution 174\AA\ coronal images from Solar Orbiter’s Extreme Ultraviolet Imager (EUI). The active region jet is simultaneously observed by the Interface Region Imaging Spectrograph (IRIS) and the Solar Dynamics Observatory (SDO).  An erupting minifilament is rooted at the edge of an active region where mixed-polarity magnetic flux is present. Minority-polarity positive flux merges and cancels with the active region’s dominant negative flux at an average rate of $10^{19}$ Mx hr$^{-1}$, building a minifilament-holding flux rope and triggering its eruption. The eruption shows a slow rise followed by a fast rise, akin to large-scale filament eruptions. EUI images and Mg II k spectra, displaying simultaneously blueshifts and redshifts at the opposite edges of the spire, indicate counterclockwise untwisting of the jet spire. This jet is the clearest, most comprehensively observed active-region jet with this instrument set, displaying striking similarities with quiet Sun and coronal hole jets. Its magnetic ($\le10^{28}$ erg), thermal ($10^{25}$ erg), and kinetic  ($10^{25}$ erg) energies suggest a significant contribution to local coronal heating. We conclude that magnetic flux cancelation builds a minifilament-carrying twisted flux rope and also eventually triggers the flux rope’s eruption that makes the coronal jet, in line with our recent results on the buildup and explosion of solar coronal jets in quiet Sun and coronal holes. That is, this active region jet clearly works the same way as the vast majority of quiet Sun and coronal hole jets.
 
\end{abstract}


\section{Introduction} 
Solar coronal jets are dynamic, transient features that appear as narrow eruptions of plasma into the solar corona. They occur in different sizes  all over the solar disk and at the solar limb.  They extend to  lengths much greater than their widths \citep{shibata11,innes16,shen21,schmieder22}. They can be found in most solar environments, including quiet Sun regions, coronal holes, and at the edges of active regions (ARs).  Some jets are accompanied with narrow coronal mass ejections (CMEs; e.g. \citealt{panesar16a,romano25}). 
Coronal jets were first observed using X-ray images from Yohkoh \citep{shibata92}, and later, they have been extensively studied with X-ray images from Hinode \citep[e.g.][]{yokoyama95,alexander99,cirtain07,savcheva07,moore18}. With the availability of higher cadence and higher spatial resolution images, coronal jets are now being  observed and analyzed using ultraviolet (UV) images and spectra from the Interface Region Imaging Spectrograph  \citep[IRIS;][]{zhangYJ21_ARjetCan,joshi21,schmieder22} and extreme ultraviolet (EUV) images from the Solar Dynamics Observatory \citep[SDO;][]{shen12,moore13,adams14}. 

\begin{figure*}[ht!]
	\centering
	\includegraphics[width=\linewidth]{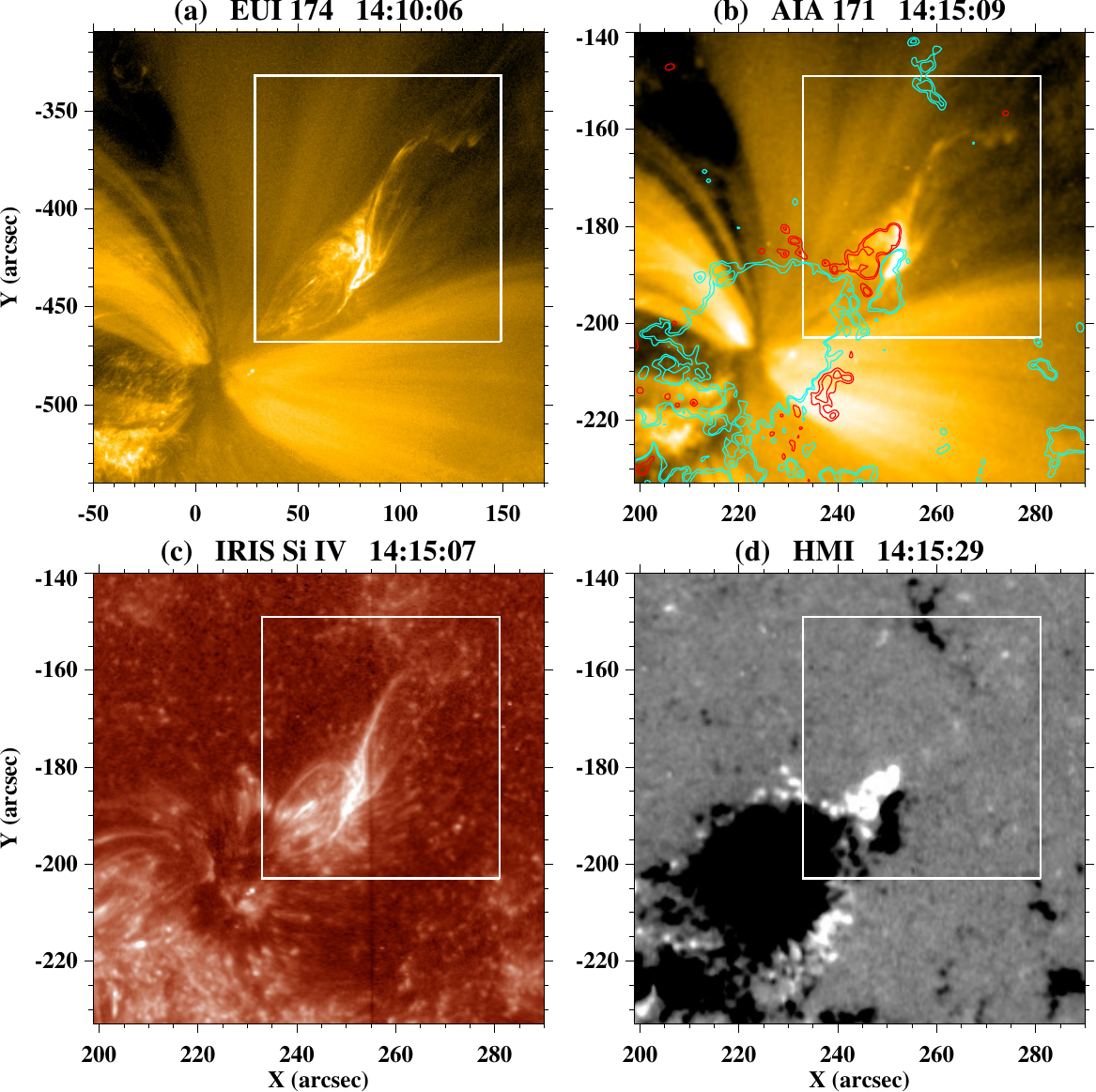}
	\caption{Active region (AR) jet observed on 29-March-2023 (NOAA 13262). Panel (a) shows an \hri\ 174 \AA\ image of the AR coronal jet. Panels (b and c), respectively,  show an SDO/AIA 171 \AA\ image and an IRIS 1400 \AA\ SJI of the same jet. Panel (d) shows an  SDO/HMI magnetogram of the AR jet-base region. The white box shows the field of view (FOV)  in Figures \ref{fig2}, \ref{fig4}, \ref{fig5}. HMI contours, of levels $\pm$50, 100 G, at 14:15:29 UT are overlaid in panel (b), where red and cyan contours outline positive and negative magnetic flux, respectively. 
		\label{fig1}}
\end{figure*}

\begin{figure*}[ht!]
	\centering
	\includegraphics[width=\linewidth]{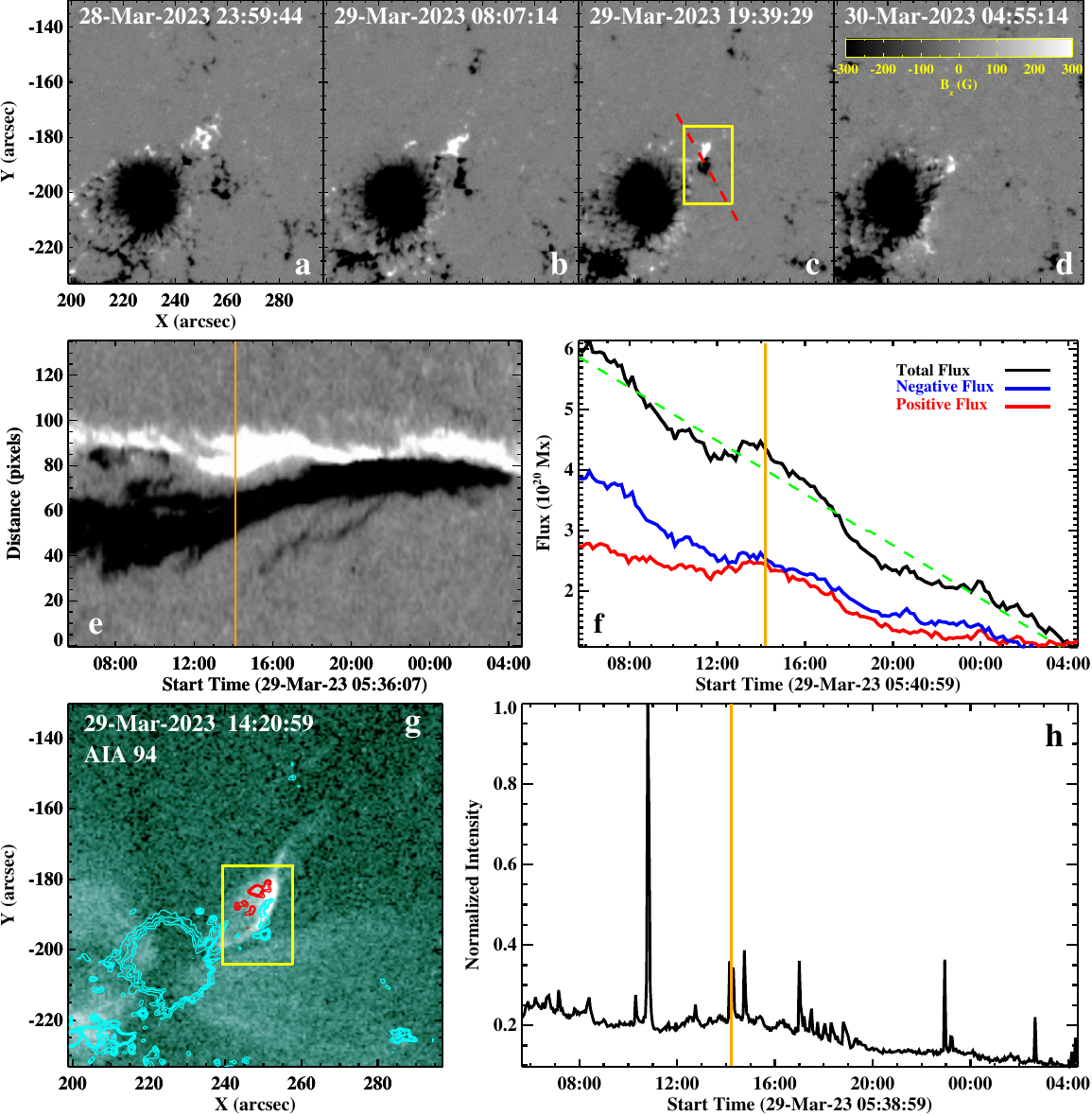}
	\caption{Evolution of active region magnetic field and AIA 94 \AA\ intensity. Panels (a--d) show the evolution of the jet-base region for two days. Panel (e) shows the time-distance map from along the red-dashed line shown in (c). Panel (f) shows the the total absolute magnetic flux (black), positive flux (red), and negative flux (blue) curves from flux that is measured inside the yellow box region of panel (c). The dashed-green line in (f) is the least square fit to the black curve. Panel (g) displays an AIA 94 \AA\ image with the main jet in the  yellow box.  Panel 
		(h) shows an intensity plot of the 94 \AA\ intensity integrated over the yellow box in panel (g). The orange vertical line marks the jet onset time.  The HMI contours, of levels $\pm$500, $\pm$400, $\pm$300, $\pm$200 G, at 29-Mar-2023 14:17:44 UT are overlaid in panel (g), where red and cyan contours outline positive and negative magnetic flux, respectively. The animation (Movie) runs from 28-Mar-2023 23:59 UT to 30-Mar-2023 04:56 UT. The animation is unannotated  and the FOV is same as shown in Panels (a--d) and (g). 
		\label{fig1a}}
\end{figure*}

Several studies using SDO EUV images have shown that coronal jets frequently originate at the locations of erupting minifilaments \citep[e.g.][]{hong11,shen12,adams14,sterling15,panesar16b,panesar18a,mazumder19_MFjetcancel,solanki19,mcglasson19,zhangYJ21_ARjetCan,mandal22}. These minifilament eruptions, which start early in the jet formation, are often triggered by magnetic flux cancelation \citep[e.g.][]{panesar16b,hong19,yang19_MFjetcancel,poisson20_ARjetCan}. A small-scale flare-like brightening, known as the jet bright point \citep[JBP;][]{sterling15},  appears under the erupting minifilament. These JBPs are evidently miniature analogs of flare arcades that form below typical coronal mass ejection-producing filament eruptions, so that coronal jet eruptions are considered scaled-down versions of such CME eruptions \citep{sterling18}. 

Studies of about 100 on-disk quiet region and coronal hole jets \citep{panesar16b,panesar17,panesar18a,mcglasson19,panesar22},  found that a large majority of jets result from minifilament (average size 10$^{4}$ km; \citealt{panesar16b}) eruptions. The pre-eruption minifilament resides in a highly sheared field magnetic flux rope above a neutral line/polarity inversion line \citep{martin86} between a majority-polarity flux patch and a merging minority-polarity flux patch. Flux convergence and cancelation at the neutral line  builds the highly sheared field and flux rope that carries the minifilament plasma \citep{panesar17}. Continuation of that flux cancelation at the neutral line eventually destabilizes the field that carries the minifilament plasma and it erupts outwards. The JBP appears over the neutral line in the eruption site. The erupting minifilament field reconnects with an adjacent oppositely-directed far-reaching magnetic field, undergoing interchange reconnection, which generates the jet spire. Reconnection-heated jet material, along with cool plasma from the minifilament, escapes along these newly reconnected far-reaching or open field lines and appears as the bright jet spire \citep{sterling15}.

The aforementioned studies demonstrate that the minifilament-eruption idea explains well a large majority of quiet region and coronal hole jets. However, active region jets often do not present as clear a picture as their quiet Sun and coronal hole counterparts \citep{sterling16, sterling17, sterling2024-ARjet}. Several factors contribute to this discrepancy: (i) the intense brightness of active regions often obscures key features of the jet eruption; (ii) minifilaments are more difficult to detect—either due to the absence of cool plasma within the flux rope, because the plasma heats rapidly making the minifilament (if any) challenging to observe, or because the early eruption of the minifilament is obscured by surrounding and/or overlying absorbing material; (iii) identifying the JBP is often uncertain; and (iv) the highly dynamic magnetic environment in active regions makes it difficult to isolate the jet-base magnetic field from the surrounding magnetic field.  Consequently, it remains unclear whether these active region jets are consistently driven 
by minifilament eruptions and whether they are consistently accompanied by flux cancelation.

Numerous other studies of active region jets exist in the literature \citep[e.g.][]{chandra15,liu-16-ARjet,mulay16,panesar16a,paraschiv20-ARjet,yang23-ARjet,garima25-ARjet,zhou25-ARjet}.  It is also important to note, however, that the primary focus of many of these studies is not on the origin of the jets, which may explain the lack of detailed investigation into these aspects. In any case, active region jets are often more complicated than quiet region and coronal hole jets. Due to their complexity, it is possible that alternative mechanisms may be at work in active region jets \citep[e.g.][]{joshi24,gou24}. Because of the difficulties with observing active region jets discussed in the previous paragraph,  determining the cause of these jets requires careful detailed analysis of individual active region jets in cases where  appropriate data are available.

Here, we present observations of an active region  jet that was fully captured by Solar Orbiter's \citep{muller2020} Extreme Ultraviolet Imager (EUI) high-resolution imager (\hri; \citealt{rochus2020}) in the 174 \AA\ passband. The \hri\ observed the jet with fast temporal cadence of 3 seconds and high spatial resolution from a pixel size of 142 km. These observations were complemented with coverage from IRIS \citep{pontieu14} and SDO \citep{pesnell12}. The jet was fully tracked by IRIS slit-jaw images and also scanned by the IRIS slit. To best of our knowledge, this paper presents the first  analysis of an active region coronal jet observed by \hri\ together with  observations from IRIS and SDO.  We will conclude that this active region jet does indeed exhibit  a  buildup and eruption process typical of quiet region and coronal hole jets.
 

\begin{figure*}[ht!]
		\centering
\includegraphics[width=\linewidth]{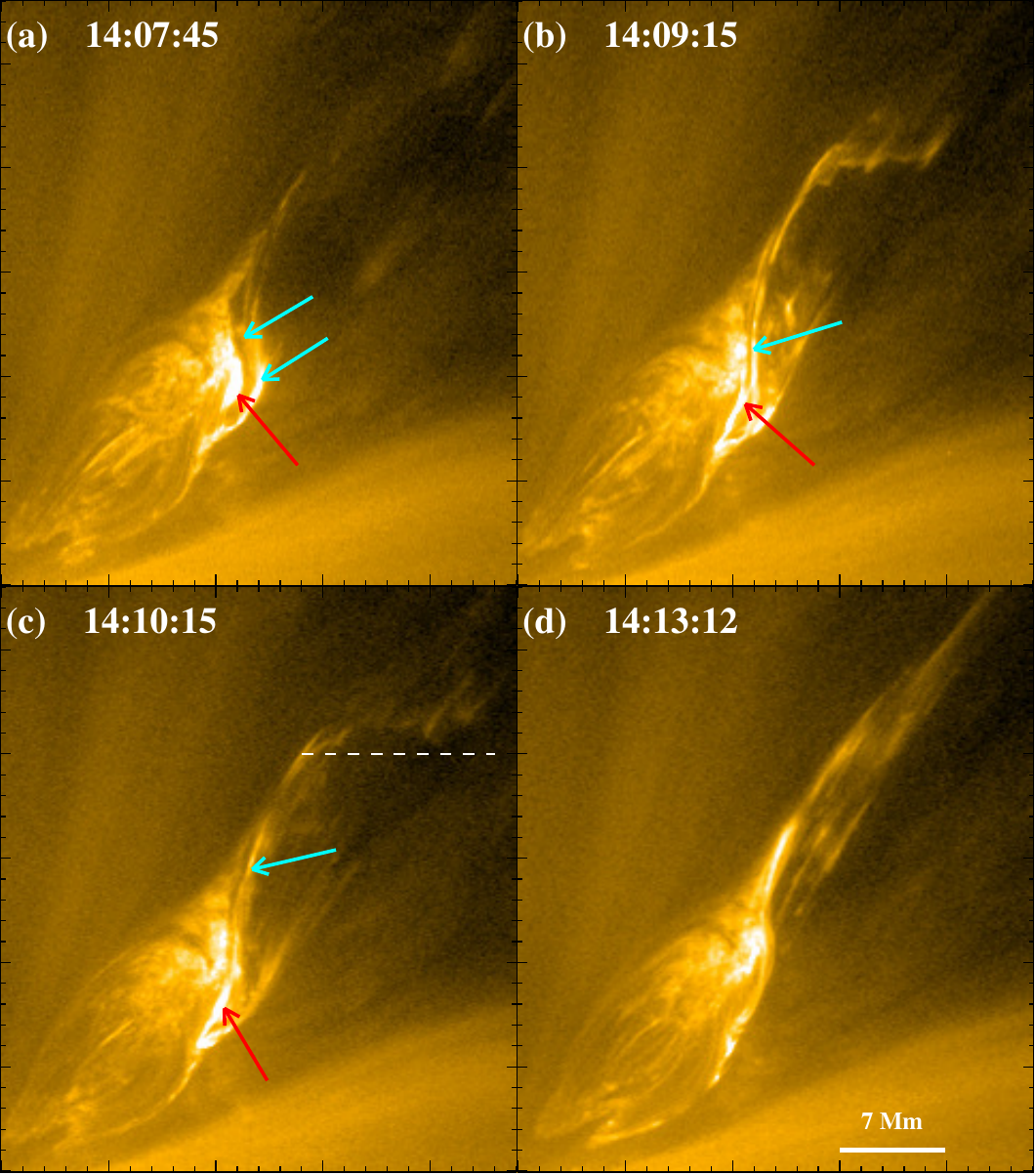}
\caption{Active region jet in \hri\ 174 \AA\ images during the eruption onset. The cyan arrows point to the erupting minifilament. The red arrows point to the jet bright point (JBP). The white dashed line in (c) shows the slice for the time-distance image in Figure \ref{fig3}. The animation (Movie1) runs from 13:40 to 14:29 UT. The animation is unannotated and the FOV is same as in this figure.
\label{fig2}}
\end{figure*}

\section{Observations} \label{sec:observations}

For our analysis, we use Solar Orbiter's \hri\ Level 2 data\footnote{https://www.sidc.be/EUI/data/L2/} from Data Release 6 \citep{euidatarelease6}. The dataset was captured on 29-March-2023.  Solar Orbiter was located at  0.394 au from the Sun.  \hri\ took images at a temporal cadence of 3 seconds with a pixel size of 142 km. \hri\ provides images in 174 \AA, emission from which is from \FeIX\ and \FeX\ lines.  The separation angle between Solar Orbiter and Earth is 2.5\degree\ on 29-Mar-2023, and solar photons reached Solar Orbiter about five minutes earlier than to SDO and IRIS. 

We utilize coordinated observations from SDO/AIA \citep{lem12}, which provides full-disk images in seven different EUV wavelengths (304, 171, 193, 211, 131, 335, and 94 \AA) with a temporal cadence of 12 seconds and a pixel size of 0.\arcsec6 (435 km). The active region jet is visible in all AIA EUV channels; however, for the analysis we primarily focus on images from the 171 \AA\ and 94 \AA\ channels. 

To examine the evolution of the photospheric magnetic field in the jet-base region, we use line-of-sight magnetograms from the SDO/Helioseismic and Magnetic Imager (HMI; \citealt{scherrer12}). The HMI magnetograms were obtained with a temporal cadence of 45 seconds, a pixel size of 0.\arcsec5 (360 km), and a noise level of approximately 7 Gauss \citep{couvidat16}. We also use the HMI vector data to display the transverse field at the jet bipole. Both AIA and HMI datasets were downloaded from the JSOC website\footnote{http://jsoc.stanford.edu/ajax/exportdata.html}. 

We also use coordinated data from IRIS on 29 March 2023. IRIS observed the active region NOAA 13262, in which our jet occurred,  approximately for four hours using its slit-jaw imager (SJIs) in Si IV 1400 \AA\ and Mg II k 2796 \AA\ filters, with a temporal cadence of 11 seconds and a pixel size of 0.\arcsec33 (240 km). A very large sparse 64-step raster was used, with a step cadence of 5.3 seconds and a raster cadence of 336 seconds, resulting in a total of 48 rasters (OBS ID 3400107460).  The spectrograph slit step size was 1\arcsec. 
We used level 2 IRIS data. The datasets had already been calibrated by subtracting dark current, flat-fielding, and correcting for geometrical distortion. For our active region jet analysis, we mainly use Si IV SJI images and rasters from the Mg II k line.

\begin{figure}
	\centering
	\includegraphics[width=\linewidth]{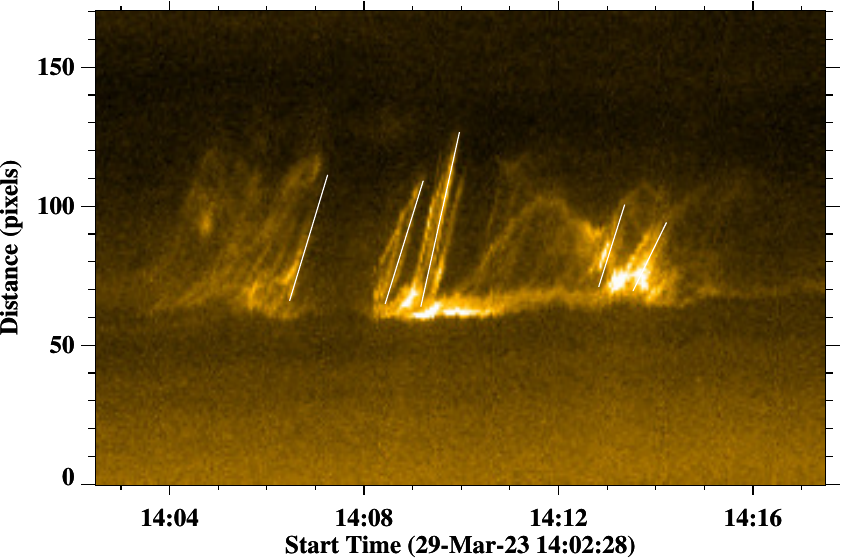}
	\caption{\hri\ 174 \AA\ time-distance map from the white-dashed line of Figure \ref{fig2}c. The plot shows the (un)twisting of the jet spire with time. The white lines trace some of the left-to-right (untwisting) motion transverse to the spire of bright strands of the spire.
		\label{fig3}}
\end{figure}

\section{Results} \label{sec:res}
\subsection{Overview}
\begin{figure*}
	\centering
	\includegraphics[width=\linewidth]{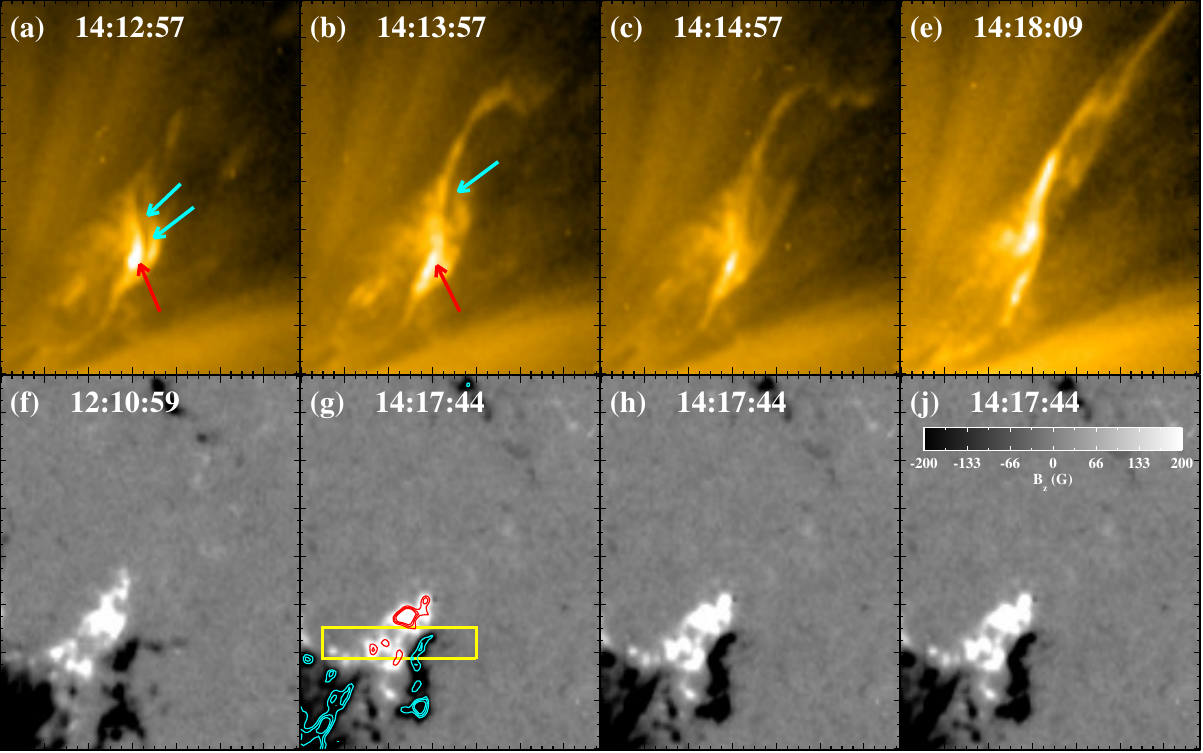}
	\caption{Active region jet in AIA 171 \AA\ images and simultaneous magnetogram. Panels (a--e) show the evolution of the erupting minifilament and jet spire. The cyan and red arrows, respectively, point to the erupting minifilament and jet base bright point. Panels (f--j) are line of sight HMI magnetograms of the jet-base region. The yellow rectangular box in (g) shows the region that is used to make the time-distance map in Figure \ref{fig4a}. The distance between major ticks is 20 pixels. The HMI contours, of levels $\pm$500, $\pm$400, $\pm$300 G, at 14:17:44 UT are overlaid in panel (g), where red and cyan contours outline positive and negative magnetic flux, respectively.
		The animation (Movie2) runs from 12:15 to 14:40 UT. The animation is unannotated and the FOV is same as this figure.
		\label{fig4}}
\end{figure*}

Figure \ref{fig1} shows an overview of active region NOAA 13262. At the north-west edge of the active region, there is a bright jet-base region (inside the white box of Figure \ref{fig1}), which produces multiple coronal jets.  Here, we mainly focus on and show a jet from the jet-base region that starts around 14:02 UT, peaks around 14:08 UT, and ends at 14:17 UT. The total duration of the jet is about 15 minutes. Because this jet is most fully captured by the \hri\ and IRIS observations, we mainly focus on this jet (hereafter `main jet'). The jet erupts from over a neutral line between the  majority-polarity (negative) flux and a patch of minority-polarity (positive) flux (Figures \ref{fig1}b,d).  The white box shows the FOV that we analyze in detail and show in Figures \ref{fig2}, \ref{fig4}, and \ref{fig5}. All the panels in these figures correspond to approximately the same time. In subsequent images (Figures \ref{fig2}, \ref{fig4}, \ref{fig5}),  cyan arrows point to the erupting minifilament, while red arrows point to the JBP.

\subsection{Recurrent Jets and Flux Cancelation}\label{sec:alljets}

Figures \ref{fig1a}(a--d) and movie shows the evolution of the active region and the jetting site. AIA data shows that there are multiple jets occurring from the same location (yellow box).  The accompanying movie shows that first these two positive and negative polarities are separated from each other and then they start to converge towards each other. Continuous flux convergence and cancelation results in multiple jets from the neutral line, ultimately leading to the disappearance of the negative flux patch from the jet base region by 04:45 UT on 30-Mar-2023.

Figure \ref{fig1a}h shows the AIA 94 \AA\ light curve that we obtained by integrating intensity over time within the yellow box region of Figure \ref{fig1a}g. The AIA 94 \AA\ lightcurve in Figure \ref{fig1a}h shows only relatively hotter events  within the box.  Every significant peak in the intensity profile exceeding the background fluctuation level corresponds to a jet originating from the same location over the course of 24 hours. As noted earlier, we focus here on the jet that is fully captured by both \hri\ and IRIS,-- the time of the main jet is marked by an orange line. Figure \ref{fig1a}g shows in the yellow box  the main jet that we will discuss in next section. It shows a faint jet spire and relatively brighter jet base bright point that lies at the neutral line. 

 Initially, the jet-base magnetic flux patches were apart from each other and no obvious jet activity was seen during this phase. Jetting began once the positive and negative flux patches started to converge and cancel at the neutral line (see MOVIE).  To illustrate the convergence of the opposite-polarity magnetic flux,  we made an HMI time-distance map (Figure \ref{fig1a}e) along the red-dashed line in Figure \ref{fig1a}c, which shows how the flux patches converge over a day. The map shows that both opposite-polarity magnetic flux patches converge toward the neutral line  and cancel with each other.  
 
 Figure \ref{fig1a}f shows the total absolute magnetic flux, along with the total negative-polarity flux, and total positive-polarity flux curves that are measured inside the yellow box region of panel (c). These curves exhibit a gradual decline over time, confirming ongoing flux cancelation at the neutral line. The total absolute flux decreases at an average rate of 2.2 $\times$ 10$^{19}$ Mx hr$^{-1}$. It should be noted that the flux patches are not fully isolated because of the continuous flows of  flux elements from the sunspot into the jet base region, but the plot presents the overall trend of the flux, which is decreasing with time. This is evidently photospheric-flow driven flux cancelation as suggested by  \cite{balle89,moore92,kaithakkal19,syntelis21,priest21,hassanin22}.

Thus, the AIA 94 \AA\ light curve reveals multiple jets originating from the same location, each of which is eventually built-up and triggered by flux cancelation at the neutral line. This behavior mimics that reported from the previous studies of quiet region and coronal hole jets \citep{panesar16b,panesar18a,muglach21}, where repeated eruptions from the same neutral line were accompanied by ongoing flux cancelation. 

\begin{figure}
	\centering
	\includegraphics[width=\linewidth]{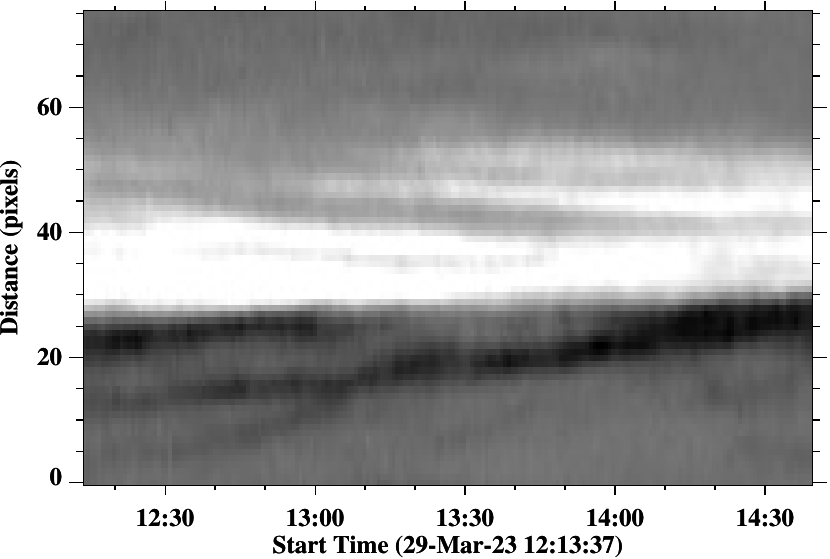}
	\caption{The jet-base magnetic flux time-distance map. The  time-distance map shows the merging of the magnetic flux inside the yellow box in Figure \ref{fig4}g. Positive and negative flux patches converge towards the neutral line with time. 
		\label{fig4a}}
\end{figure}

\begin{figure*}
	\centering
	\includegraphics[width=\linewidth]{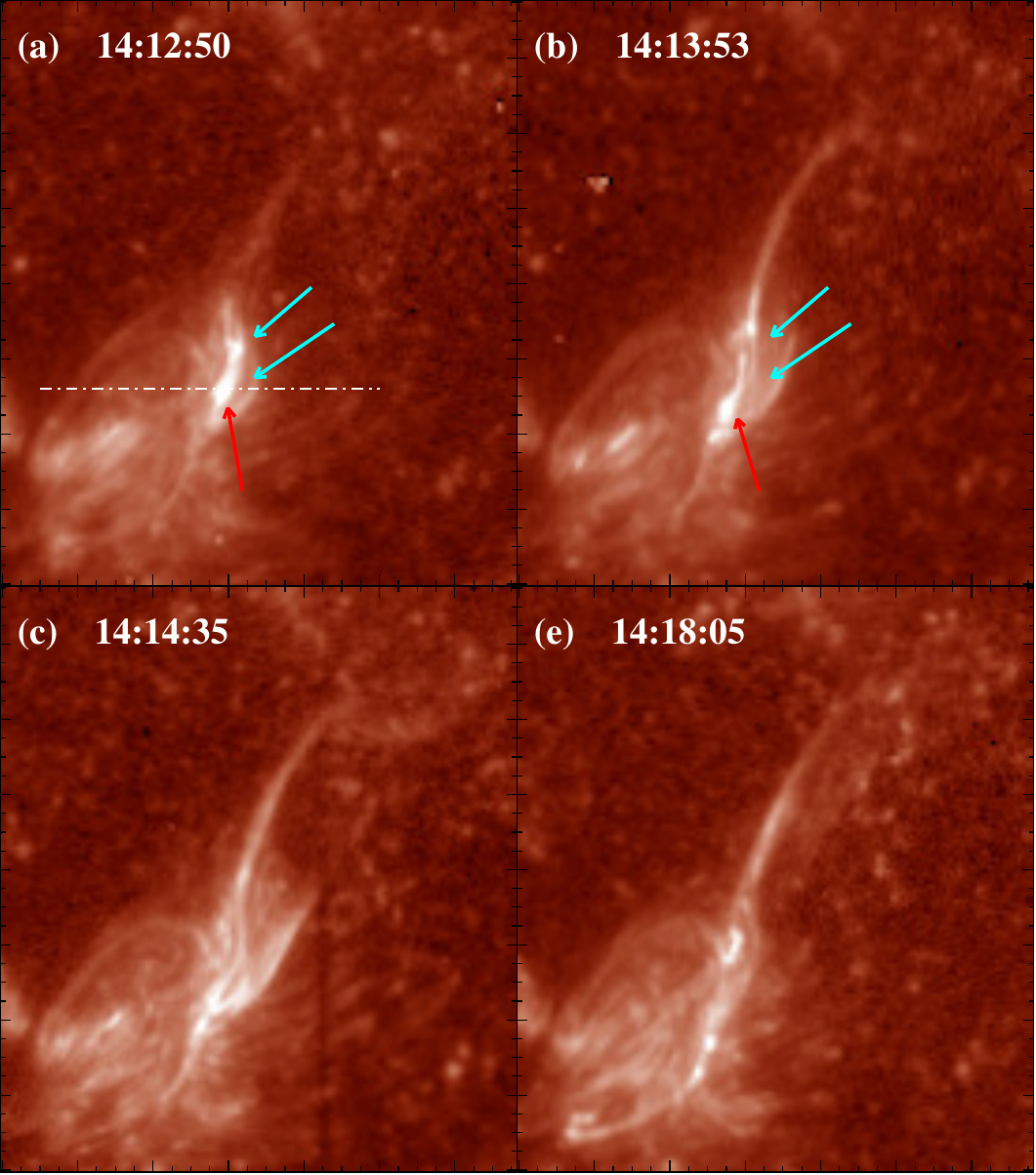}
	\caption{Active region jet observed by IRIS in 1400 \AA\ slit-jaw images. The cyan and red arrows, respectively, show the erupting minifilament and jet base bright point. The dashed dotted line in (a) marks the slice for the time-distance map of Figure \ref{fig5a}. The distance between major ticks is 20 pixels. The animation (Movie3) runs from 12:40 to 14:40 UT. The animation is annotated same as this figure and the FOV is same as this figure. 
		\label{fig5}}
\end{figure*}

\subsection{Main Jet: \hri, SDO, and IRIS Observations}\label{sec:mainjet}

Figure \ref{fig2} shows \hri\ images of the main jet and the accompanying animation shows the complete evolution of the jetting region for about one hour of time period. The AIA 94 \AA\ lightcurve shows a peak in the intensity curve around the time of the orange line (Figure \ref{fig1a}h) during this jet. The jet starts at 14:02 UT and continues for about 15 minutes.
It is apparent from the \hri\ images that there is a minifilament (pointed to by cyan arrows) that is present in the jet base region and is starting  to slowly lift off at 14:01:21 (see Movie1) -- this is a earlier strand of the minifilament (see \citealt{sterling17} paper for minifilament strand details) erupting earlier than the bulk of the body of the minifilament, and that  earlier stand's eruption  leads to the first burst of spire material around 14:03 UT. At that time, a small-scale brightening starts under the erupting minifilament.  The main (or the main strand of the) minifilament plasma vigorously erupts at 14:07:00 UT  as the JBP grows underneath the erupting minifilament (Figure \ref{fig2}a). The JBP is also visible in hotter AIA channels, for example in AIA 94 \AA\ (Figure \ref{fig1a}g). Intermittent peeling of cool plasma occurs on the west side of the jet base, as flux ropes or threads erupt  with  cool plasma. The minifilament then enters a vigorous spinning phase from about 14:08:24.  This thus appears to be a fast rise,  occurring at the time of the start of the (un)twisting.  Thus, the untwisting onset and the fast-rise onset are about concurrent. The \hri\ images clearly show the ejection of hot and cool material in the jet spire, which forms on the side of the erupting minifilament away from the internal reconnection, consistent with the explanation and schematic in Figure 2 of \cite{sterling15}.  The jet spire ejects outwards with an average speed of 250 $\pm$ 15 \kms\ between 14:12:30 UT and 14:14:30 UT.



The \hri\ animation shows a clear indication of rotation (spinning) of the jet spire. 
If we view the spire from its top, this apparent spire rotation is  counterclockwise about its axis. The IRIS raster data verifies the counterclockwise spinning of the jet spire, which is discussed in detail in Section \ref{sec:iris}. To present additional evidence  of rotation of the jet spire, we took a cut across  the spinning jet spire (Figure \ref{fig2}c) to obtain a time-distance map. Figure \ref{fig3} shows the \hri\ time-distance map for the untwisting motion of bright strands of the jet spire. Some of the strand motions (transverse to the spire) are tracked and highlighted with white lines. The average transverse speed of these tracked strands is 138 $\pm$ 37 \kms. Similar transverse motions are also seen along a `spine' by \cite{petrova24} using EUI data. We find that between 14:10 and 14:13 UT (Figure \ref{fig3}) the spire gets narrower, producing the downward (negative) slope. While we cannot rule out that some other dynamical process may result in the Figure \ref{fig3} trajectories, these observations are consistent with our suggestions of an erupting twisted flux rope resulting in what we interpret from the data as evidence for an untwisting jet spire. We will present further evidence based on magnetic field data for our viewpoint in Section \ref{sec:pfss}.

\begin{figure}
	\centering
	\includegraphics[width=\linewidth]{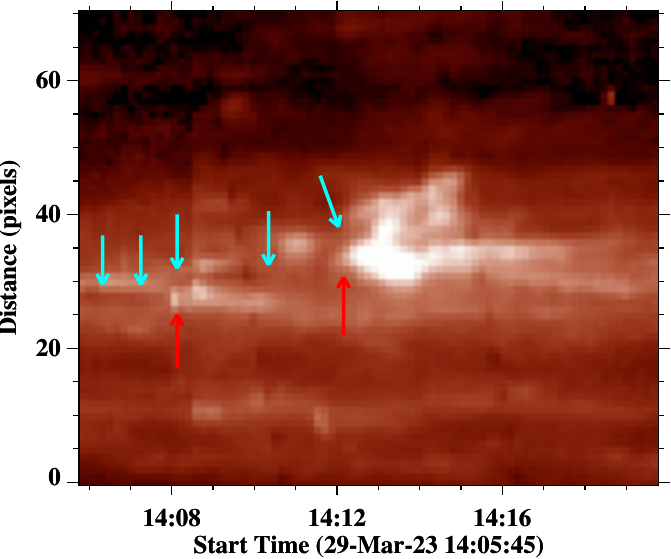}
	\caption{Evolution of the minifilament in Si IV. This is a time-distance map plotted along the dotted-dashed line of Figure \ref{fig5}a. The cyan and red arrows, respectively, point to the minifilament during slow-rise and fast-rise phase, and jet base bright point brightening.
		\label{fig5a}}
\end{figure}

Figure \ref{fig4} and the accompanying animation display the same jet in AIA 171 \AA\ images, along with photospheric line-of-sight HMI magnetograms of the jet’s base region. These images are at approximately the same time as the \hri\ images shown in Figure \ref{fig2}. Red arrows point to the JBP that forms beneath the erupting minifilament. While the main features of the jet -- such as the jet bright point JBP and the minifilament -- are visible in the AIA 171 \AA\ images (we inspected other AIA EUV channels as well, including 304 \AA, finding jet features much clearer in \hri\ than in any of the  AIA channels), the distinct spin of the jet spire is much clearer in the high-resolution \hri\ images and is not as apparent in AIA.  This difference highlights the importance of high spatial resolution in detecting fine-scale dynamics \citep{sterling23}, such as  untwisting motions, which are consequences of the magnetic-reconnection  mechanism making coronal jet spires.  


The jet occurs above the magnetic neutral line, located between the majority-polarity (negative) flux and an intruding minority-polarity (positive) flux. The JBP brightens on  this neutral line, precisely where the minifilament was anchored prior to its eruption (Figures \ref{fig1a}g and \ref{fig4}). As discussed earlier (Section \ref{sec:alljets}), all jets in this region are built-up and triggered by continuous flux cancelation at the neutral line. Figure \ref{fig4a} presents an HMI time–distance map generated by summing the signed flux along the y-direction within the yellow box shown in Figure \ref{fig4}g. This region was chosen to encompass the neutral line where the flux cancelation is taking place. The map displays the convergence and cancelation of magnetic flux occurring before, during, and after the main jet event. The base of the jet is against the northwest side of the negative-polarity lone leading sunspot of the jet's active region (NOAA 13262).  This active region is inside a larger region of negative-polarity magnetic network.  The positive flux clump in the jet's base is surrounded by negative flux but it does not have a true magnetic null above it.  Instead, component reconnection occurs low in the corona above the positive flux clump (see details in Section \ref{sec:pfss}). The pre-jet flux cancelation at the neutral line in the jet's base  is appropriate for building the minifilament flux rope and triggering its eruption.  The minifilament eruption is appropriate for driving interchange (breakout \citealt{antiochos98}) reconnection, thereby making the jet spire.  This mimics our previous quiet region and coronal hole jet and jetlet observations \citep{panesar16b,panesar18a,panesar18b,panesar19}.

\begin{figure*}
	\centering
	\includegraphics[width=\linewidth]{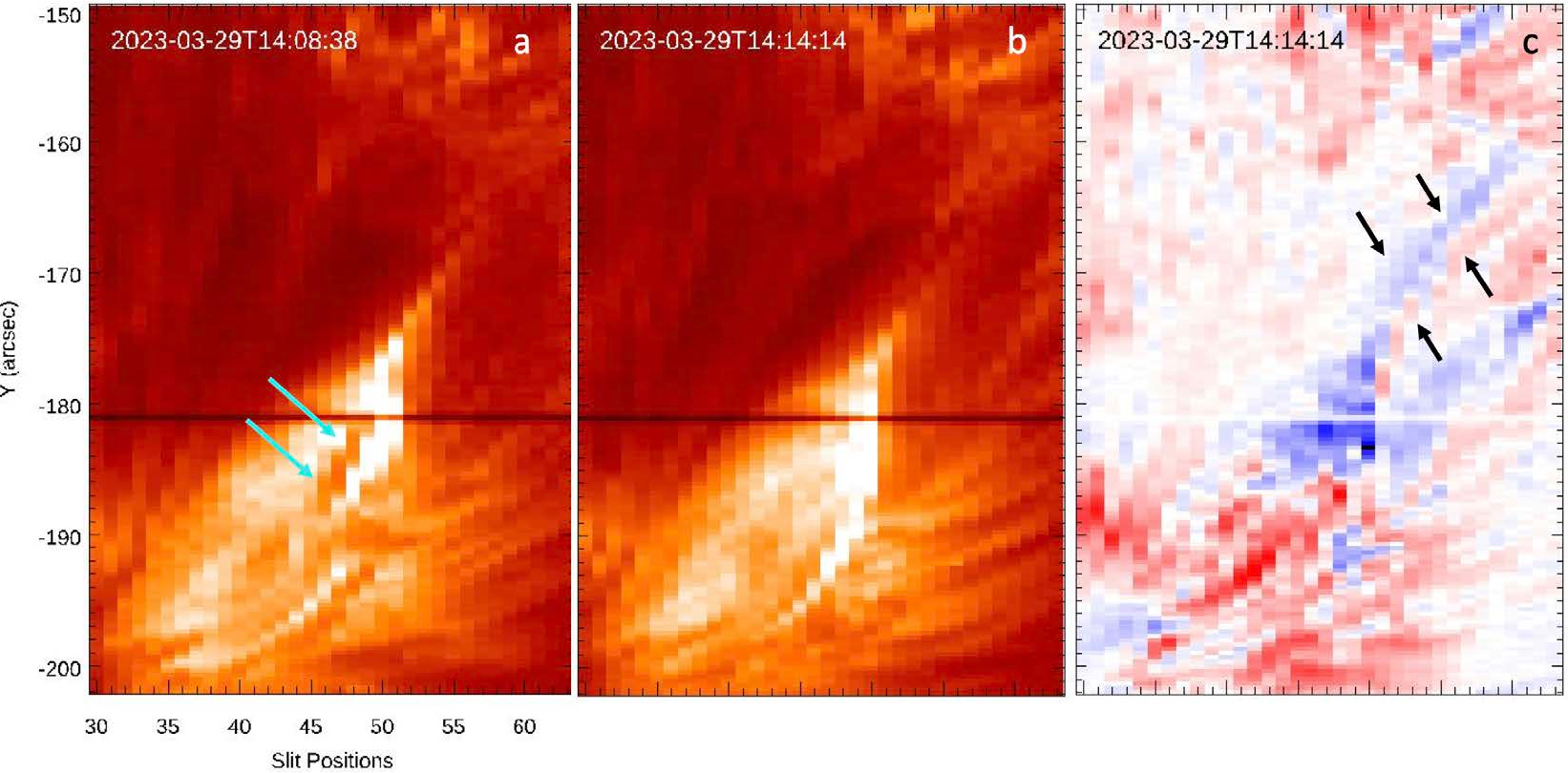}
	\caption{Jet onset. Panels (a) and (b) each show a spectroheliogram of the Mg II k line during the jet eruption. The cyan arrows point to the erupting minifilament. Panel (c) shows the Dopplergram of the jet onset. The upper and lower black arrows, respectively,  point to the blueshifts and redshifts along the jet spire. 
		\label{fig6a}}
\end{figure*}

Figure \ref{fig5} and the accompanying movie show the full evolution of the jet as observed in the IRIS SJI 1400 \AA\ filter. The cyan arrows point to the minifilament as it begins to rise, while the red arrows  point to  the  brightening (JBP) that appears beneath the erupting minifilament. This behavior is consistent with our \hri\ observations shown in Figure \ref{fig2}. As the minifilament continues to erupt, the jet spire forms.

Near the start of the eruption, the IRIS Si IV SJI images show the minifilament structure in the jet base region. To capture the temporal evolution of the minifilament, we generated a time–distance map along the dotted-dashed line in Figure \ref{fig5}a. The Si IV time–distance map (Figure \ref{fig5a}) reveals that the minifilament remains nearly stationary at first (see the first two leftmost cyan arrows). Around 14:08 UT (as a reminder \hri\ sees this 5 minutes ahead), a small-scale brightening (first red arrow) appears beneath the minifilament, marking the beginning of  the eruption's slow-rise phase. This reconnection-caused brightening accompanies a noticeable upward motion of the minifilament, as seen in Figure \ref{fig5a} (the right most cyan arrow). 

The minifilament eruption remains in this slow-rise phase for approximately four minutes. Then, at around 14:12 UT, it abruptly transitions into its fast-rise phase, entering its full eruption, between the last two rightmost cyan arrows. During the fast-rise phase, a prominent JBP forms in the wake of the erupting minifilament (see last red arrow). Similar two-phase eruptions—beginning with a slow rise followed by a rapid eruption—have also been reported in on-disk, quiet-Sun pre-jet minifilaments \citep{panesar20b} and active region erupting minifilaments \citep{zhang21}. The two-phase minifilament eruption is another evidence that these eruptions are miniature versions of typical larger solar filament eruptions because larger filament eruptions often show such slow- and fast-rise phases during the eruption \citep{sterling05,mccauley15}. 

It is worth noting that the distinct slow- and fast-rise phases of the minifilament are not clearly visible in the \hri\ and AIA 171 \AA\ images, likely due to the presence of foreground coronal emission that obscures the view.

\subsection{Mg II k Observations of the Main Jet}\label{sec:iris}

The active region jet was observed by the IRIS slit raster with a cadence of 5.6 minutes, enabling spectroscopic analysis of the event. From each raster scan, we constructed a spectroheliogram at the core of the Mg II k line (2796.38 \AA), which corresponds to chromospheric heights. This allows us to examine the spatial distribution and temporal evolution of chromospheric features of the jet.

Figure \ref{fig6a} presents two Mg II k spectroheliograms taken at different times: the first (Figure \ref{fig6a}a) just before the appearance of the jet spire, and the second (Figure \ref{fig6a}b) during the onset of the spire. The erupting minifilament is clearly visible in the first spectroheliogram. However, due to the relatively coarse raster cadence, we were unable to follow the  temporal evolution of the minifilament in much detail. The observed structure closely resembles minifilaments seen in coronal hole jets in IRIS Mg II k spectroheliogram \citep{panesar22}, and is also similar to features reported in H$\alpha$ by \cite{hermans86} and \cite{wang00}.

Figure \ref{fig6a}c shows the Dopplergram derived from the Mg II k spectroheliogram in Figure \ref{fig6a}b, constructed using the intensity difference at $\pm$15 km s$^{-1}$. The Dopplergram reveals opposite Doppler shifts on the two sides of the jet spire, providing evidence of spinning motion. When viewed head-on to the spire, the observed Doppler-shift pattern corresponds to counterclockwise rotation of the spire about its axis: the blueshifted  is on the north side, and the redshifted is on the south side of the spire. Similar Doppler signatures—indicative of spinning of the spire using IRIS data have been reported by \cite{cheung15} in active region jets, by \cite{panesar22} in coronal hole jets,  by \cite{tiwari18} in penumbral jets, and by \cite{kayshap21} in surges. 

A caveat is that a similar Doppler pattern might also result from some other factors, such as  plasma flows in a static, helical geometry field without any untwisting motions in the magnetic field itself. While we cannot rule out all such possibilities, the untwisting concept matches what we appear to observe in the 174 \AA\ movie, and it matches our expectation for the appearance of the time-distance plot of Figure \ref{fig3}.  Therefore, we conclude that the untwisting motion is the most plausible explanation for the observed Doppler pattern in Figure \ref{fig6a}c.



\begin{figure*}
	\centering
	\includegraphics[width=\linewidth]{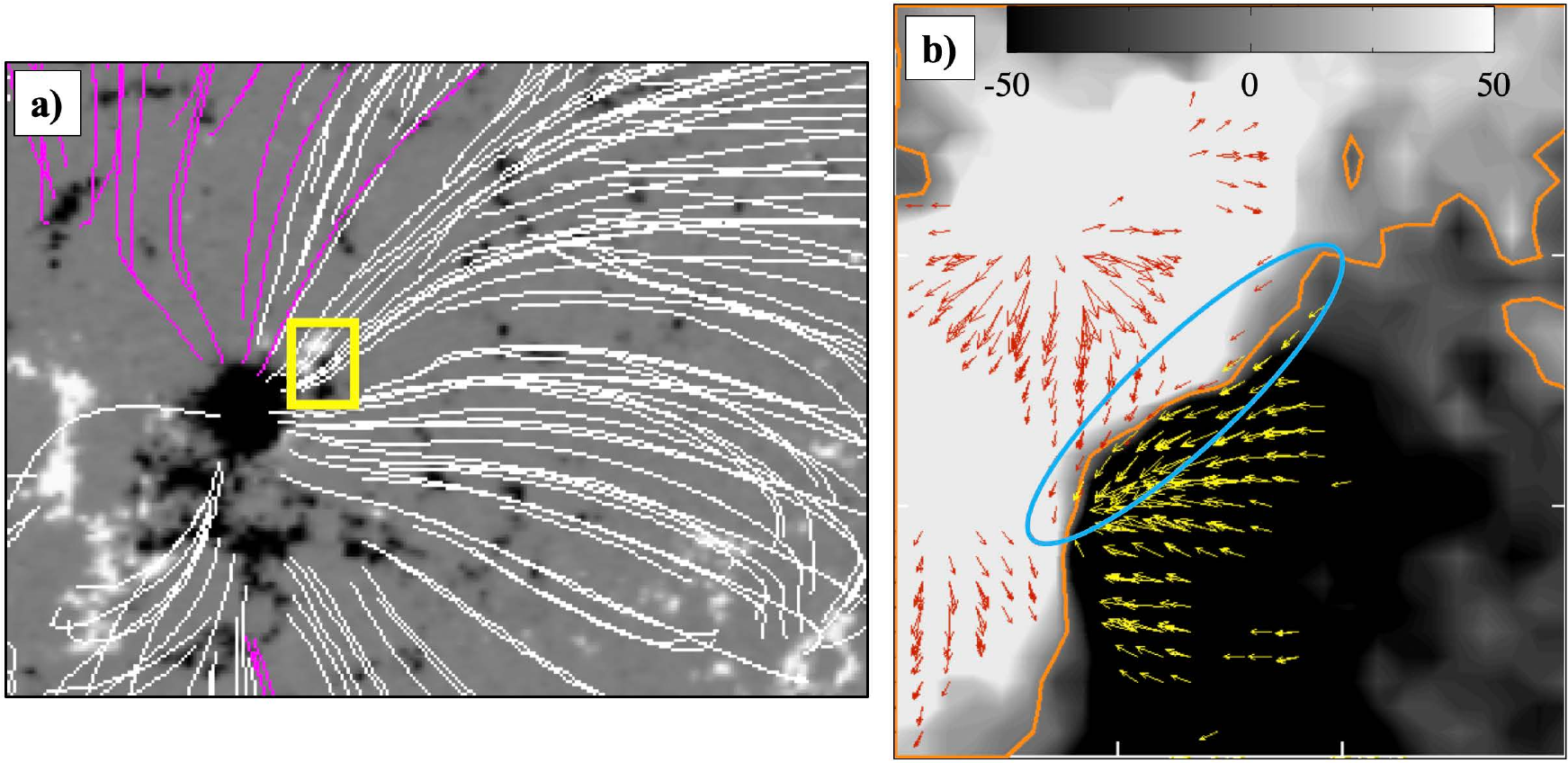}
	\caption{Magnetic field topology at the jet base. Panel (a) shows the Potential Field Source Surface (PFSS) extrapolation of the jet and surrounding regions over-plotted on a Bz map at 14:00 UT. The pink and white color  represent, respectively, open and close field lines. The yellow box shows the field of view displayed in (b). Panel (b) shows the transverse field vectors over-plotted on an HMI Bz map at 14:00 UT. The orange contour between the positive and negative flux of the jet bipole is the magnetic neutral line. The red and yellow arrows display the transverse magnetic field vectors of the positive and negative flux clumps, respectively. The blue oval outlines the region showing left-handed shear.
		\label{pfss}}
\end{figure*}

\begin{figure*}
	\centering
	\includegraphics[width=\linewidth]{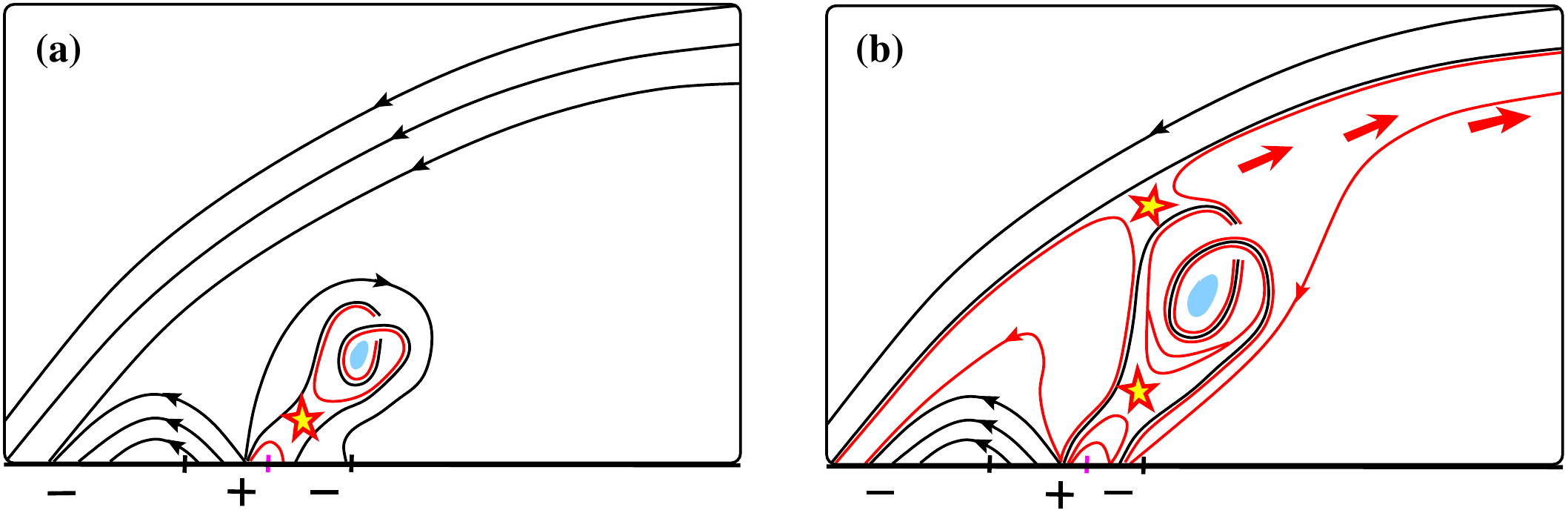}
	\caption{Schematic depictions of the minifilament flux rope eruption that makes the jet's base brightening and the jet's spire based on our observations and PFSS extrapolations. The drawings are schematics of the erupting and reconnecting magnetic field viewed horizontally from the south along the direction of the neutral line between the positive flux in the jet's base and the negative flux clump that is canceling with that positive flux.  The left-side negative flux in the schematic represents negative flux near the sunspot's northwest edge in the HMI magnetograms. Panel (a) shows the onset of fast internal reconnection. Panel (b) is during the onset of fast external reconnection. The thick horizontal black line is the solar surface. Positive and negative magnetic flux polarities are labeled with ``+" and ``–" signs, respectively.  Stars indicate the locations of ongoing magnetic reconnection. The curves are projections of magnetic field lines on a vertical east-west plane through the middle of the jet's base. Red curves are projections of field lines that have undergone reconnection.  Black curves are projections of field lines that have not yet undergone reconnection or that will not undergo reconnection.  The sky blue cloud carried in the flux rope's spiral field is the dark filament viewed end-on.  The drawings are for the spiral field having left-handed twist.  The magenta vertical tick marks the magnetic neutral line, and the black vertical ticks mark the outer boundaries of  ``+" and ``–"  flux. In panel (b), the solid red arrows represent plasma outflow from the reconnection out along the reconnected far-reaching field that fills the jet spire. 
		\label{skt}}
\end{figure*}

\subsection{Jet energies}

We estimate an upper bound on the magnetic energy released during the active region jet using the formula B$^{2}$ $\times$ V/8$\pi$, resulting in a value of approximately 4.7 $\times$ 10$^{28}$ erg, where B is the magnetic field strength and V is the volume of the pre-eruption magnetic field. 
For this calculation, we take a magnetic field strength of 300 G near the neutral line. This value of the field strength is based is on the measurement of the magnetic field strength near the magnetic neutral line (see magnetic field contours in Figures \ref{fig1a}, \ref{fig4}). This estimate of the magnetic field is consistent with that found by \cite{sterling17} in another active region with a somewhat similar magnetic field set-up. We note that the magnetic field strength can vary significantly within jet-producing regions across different parts of an active region. In particular, values of 500 G and above have been inferred in low-coronal flux ropes within active regions using nonlinear force-free field  extrapolations \citep{thalmann14}.

For estimating the volume (V), we assumed a cylindrical geometry (V = $\pi$r$^{2}$l), and took the length of the minifilament as l and its width as twice the radius 2r \citep{sterling17}. The average length of our minifilament is 12\arcsec\ (the average minifilament length is also verified from the neutral line's length from HMI magnetogram) and the width of the minifilament is 2\arcsec.  
 It is important to note that accurately determining the length and width of the minifilament, and of the magnetic flux rope in which the minifilament resides, is challenging due to limitations in spatial resolution and projection effects. As a result, the values used in this study represent approximate measurements based on visual inspection of the available data.  Our active region jet magnetic energy  estimate is comparable to that of active region  jets \citep{sterling17}, braided active region loops \citep{cirtain13}, and subflares \citep{tiwari14}. However it is about one or more orders of magnitude higher than the magnetic energy of small-scale phenomena such as some campfires \citep{panesar2021}, some coronal hole jets \citep{pucci13},  some coronal bright points \citep{priest94}, and some bright dots \citep{tiwari19,tiwari22}. 

We also estimate the thermal and kinetic energy of our active region jet. The thermal energy is given by E = 1.5N$_{e}$k$_{B}$TV, and the kinetic energy by 0.5N$_{e}$m$_{p}$V$v^2$, where N${e}$ is the electron number density, k$_{B}$ is the Boltzmann constant, T is the plasma temperature, V is the volume, m$_{p}$ is the proton mass, and $v$ is the  jet spire speed. Using representative values—$v$ = 250 km s$^{-1}$, N$_{e}$ = 10$^{10}$ cm$^{-3}$, and T = 10$^7$ K—we estimate the thermal energy to be on the order of 1.9 $\times$10$^{25}$ erg and the kinetic energy to be around 6.8 $\times$10$^{25}$ erg.  The estimated kinetic energy is consistent with values reported for coronal microjets \citep{hou21} and they are higher than the values (10$^{21}$ erg) reported by \cite{chitta23} for picoflare jets. 
Additionally, our coronal jet transiently heats the corona locally with a thermal energy of order 10$^{25}$ erg.  This is much more than the energy (of order 10$^{24}$ erg) estimated by \cite{parker88} for nanoflares.  But if nanoflares exist in sufficient quantity and frequency throughout the corona, then similar to our active region jet they might locally and transiently supply thermal energy to the corona, but at a sufficient rate and sufficient spatial distribution to account for the observed coronal emissions, as suggested by \cite{parker88}.

To compare energy estimates with  quiet region jets, we computed the upper-bound magnetic energy of quiet region jets \citep{panesar16b}, which comes out to be around  10$^{28}$ erg. Furthermore, the thermal and kinetic energies are both on the order of 10$^{25}$ erg for quiet region jets.  
Given their energy content, particularly in the magnetic component, these jets are capable of releasing enough energy to heat the quiet Sun corona locally, supporting the idea that small-scale dynamic events  in quiet regions may power coronal heating (\citealt{moore11}  suggest that Type II spicules might fulfill this role.)

\subsection{Magnetic Set-up of the Jet}\label{sec:pfss}

Figure \ref{pfss}a shows the Potential Field Source Surface (PFSS) extrapolation \citep{schrijver2003} of the jet site. It shows that open and/or far-reaching magnetic field lines are rooted at the edge of the sunspot.  These open/far-reaching field lines overlie the jet bipole. Presumably, the  erupting minifilament field reconnects with the  open/far-reaching field lines via component reconnection and and produces the jet spire. 
	
	We also examined the HMI vector data to see the direction of the transverse field at the jet neutral line (Figure \ref{pfss}b). The direction of the vectors (inside the blue oval) show that the magnetic field is sheared at the neutral line and that the shear is left-handed. For the potential field to acquire left-handed shear near the neutral line through shear flows along it, the positive flux would need to move northwest and the negative flux southeast along the neutral line. This flow pattern imparts left-handed shear to the field at the neutral line, with the sheared field pointing horizontally toward the southeast. In Figure \ref{pfss}b, the vectors closest to the neutral line clearly show this left-handed shear. Such negative shear implies that the flux rope carries left-handed twist. That is, flux cancelation at the neutral line builds a left-handed twisted flux rope, consistent with the model of \citet{balle89}.  In Figure \ref{fig1}b, the jet's JBP flare arch straddles the neutral line's sheared-field interval, consistent with our scenario for this jet eruption. 

	
Figure \ref{skt} is a schematic of the minifilament flux rope eruption that is  consistent with our observations. 
The minifilament sits in a flux rope above the neutral line in the  highly sheared field. The cool minifilament plasma is suspended within the twisted flux rope that is embedded in the sheared core field above the jet base neutral line.  The core field exhibits left-handed shear (Figure \ref{pfss}b).  The continuous flux cancelation at the neutral line (1) builds the flux rope that has left-handed twist, and then (2) destabilizes the field holding the minifilament plasma and it erupts outwards (Figure \ref{skt}a). First the outer envelope of the erupting flux rope and then the flux rope reconnect via component interchange reconnection (upper star in Figure \ref{skt}b) with overlying canopy field (in our case the canopy from the negative flux of the sunspot).  Component reconnection happens when the crossing angle between two field lines is beyond a sufficiently large (critical) value \citep{parker88,klimchuk2015,klimchuk2023,moore24}. According to \cite{parker88} the critical crossing angle should be about 15$\degree$. Figure \ref{skt}b is during the time of both continuing fast internal reconnection under the erupting flux rope and fast external reconnection of the flux rope's field with encountered overlying canopy field from negative flux near the sunspot's northwest edge.  The reconnected far-reaching field in the jet spire initially has left-handed twist in its lower legs and feet.  That twist escapes (untwists) out into the far reaches of the reconnected field.  The untwisting of the left-handed twist gives the spire's field (viewed from in front of the oncoming spire outflow) counterclockwise rotation about the interior of the spire \citep{moore15}. 

\section{Discussion}

We investigated in detail the evolution and dynamics of an active region jet observed by \hri\  with exceptionally high temporal cadence (3 seconds) and fine spatial resolution
 (pixel size of 142 km). The jet was simultaneously observed by IRIS in  Si IV
  slit-jaw images (SJI), enabling us to study its characteristics in the transition
   region, complemented by raster spectra in the Mg II k line. The photospheric
   magnetic field evolution at the jet base was examined using line-of-sight magnetograms from SDO/HMI, co-aligned with SDO/AIA 171 \AA\ images.

The high-resolution \hri\ data revealed fine-scale structures within the jet that are not discernible in the lower-resolution AIA images. This event serves as an exceptionally well-observed example of an active region jet,  in comparison with previously reported active region jets \citep[e.g.,][]{sterling16, mulay16,sterling17}. Notably, the base of the jet is a distinct and isolated bipole, minimally influenced by nearby emerging flux regions, making it ideal for detailed analysis.
Often in active region jets the jet-base regions are not isolated from other rapidly evolving magnetic field, which makes it difficult to determine their cause and triggering mechanism. Furthermore, for many active region jets the  bright emission from the surrounding active region, and/or surrounding absorbing material \citep{sterling2024} also makes it difficult to discern whether there is an erupting  minifilament. 

In our previous work, we have developed a picture for how we believe many coronal jets work: flux cancelation results in a minifilament that erupts to make the jet.  We have also previously argued that a twist on that erupting minifilament can transmit twist to the coronal field through magnetic reconnection, and the unraveling of that twist can result in the twisting of the jet’s spire. We developed that picture mainly from extensive observations of quiet Sun and coronal hole jets. To date, no research has looked in the detail presented here at an active region jet with a spinning spire, from the standpoint of that erupting-minifilament idea. Here, we argue that the flux-cancelation/minifilament-eruption scenario plausibly explains our present observations.  

We followed the evolution of the jet base region for two days and noticed that the jetting site (the jet base bipole)  produced multiple jets within the time period of two days. The foot of a newly emerged minority  flux (positive flux) patch near the sunspot starts to cancel with a nearby pre-existing negative flux clump. The flux cancelation between the newly emerged positive flux patch and the pre-existing negative flux results in the multiple active region jets. The jetting activity ceases once the negative flux is fully canceled at the neutral line. With the disappearance of the neutral line, the magnetic conditions necessary for jet formation apparently no longer exist, and no further jets are produced. All of these jet eruptions show-up well in the AIA 94 \AA\ light curve during the continuous flux cancelation at the neutral line. Thus, our active region jets are consistent with quiet region jets, which have also been observed to recur multiple times during flux cancelation between newly emerged minority-polarity flux and pre-existing majority-polarity flux patches \citep{panesar17}. This behavior is also seen in  studies of individual jets, where each jet event is evidently triggered by flux cancelation at the neutral line \citep[e.g.][]{hong11,adams14,young14a,young14b}. Our example jet  supports that magnetic reconnection in convection-driven  flux cancelation is the fundamental process responsible for the formation and triggering of jets. 

During times in our observation period when flux is emerging with no flux cancelation, no obvious typical coronal jets occur. Some weak jets erupt from neighboring neutral lines but not from the emerging-flux  neutral line.  In our jet, the overall total magnetic flux canceled with an average rate of 2.2 $\times$ 10$^{19}$ Mx hr$^{-1}$,  in similar order to that of other active region jet flux cancelation rates \citep{sterling17} as well as similar to the pre-eruption flux cancelation rates of CME-producing filament eruptions  \citep[10$^{19}$ Mx hr$^{-1}$][]{sterling18}. The observed cancelation rate is an order higher than that of quiet region jets \citep[10$^{18}$ Mx hr$^{-1}$;][]{panesar16b}, coronal jets \citep[10$^{18}$ Mx hr$^{-1}$;][]{panesar18a}, jetlets  \citep[10$^{18}$ Mx hr$^{-1}$;][]{panesar18b}, and campfires  \citep[10$^{18}$ Mx hr$^{-1}$;][]{panesar2021}.

We mainly focus on a  jet (`main jet') that was fully captured by \hri, IRIS and SDO datasets. Here we summarize the  main findings from this event: (i) A minifilament  was observed at the neutral line (12\arcsec\ long), at the site of ongoing magnetic flux cancelation; (ii)  IRIS Si IV slit-jaw images revealed a two-phase eruption: an initial slow-rise phase followed by a fast-rise phase during the eruption onset analogous to larger-scale filament eruptions; (iii) the jet spire, as seen in \hri\ images, indicate counterclockwise untwisting while extending outwards; (iv) Dopplergrams from the Mg II k line indicated the presence of both blueshifted and redshifted plasma along the jet spire, strongly supporting that the EUV-inferred motions were due to a physical twisting of the jet's spire, and  that the untwisting motion was counterclockwise.

The lifetime of this active region jet is 15 minutes, which is similar to durations reported in previous observations of coronal jets \citep[e.g.][]{mulay16,panesar16a,sterling17}. The jet spire extends outward with an average speed of  250 $\pm$ 15 \kms,  in the speed range of active region jets reported in earlier studies  \citep{panesar16a,zhang23-jethunter,musset24}. The length of our pre-jet minifilament (12\arcsec) is comparable to the lengths of pre-jet minifilaments observed in both active and quiet region jets \citep{panesar16b,sterling17}. However, its width, about 2\arcsec\, is narrower than those typically found in quiet region pre-jet minifilaments  \citep{panesar16b} but it is similar to the size of the ``minifilament strands" seen in the active region jets of \cite{sterling17}. This is in accord with larger solar filaments, in that quiescent filaments are typically wider than active regions filaments \citep{mackay10}. 

Small-scale flux ropes seen as minifilaments are commonly observed throughout the solar atmosphere. They are not only present at the bases of coronal jets but are also found in coronal bright points that appear at sites of flux cancelation, found in both observations and simulations \citep[e.g.][]{madjarska22,nobrega-siverio22}, as well as in campfires \citep[e.g.][]{panesar2021}, and quiet region upflows \citep[e.g.][]{schwanitz23}. In addition, small-scale flux ropes have been identified in simulations—such as those by \cite{chen21}, which show them forming at the base of campfires, and in Bifrost simulations of fine-scale jets \citep{panesar23}.

The \hri\ images show a clear indication of counterclockwise spin of the jet spire.  
The Mg II k spectroheliograms show the presence of  Doppler shift along the jet spire, confirming counterclockwise untwisting motion of the magnetic field in the jet spire.  Such opposite Doppler shifts next to each other in jet spires is reported by, e.g., \citealt{cheung15}, for active region jets and by \cite{panesar22} for quiet region jets using IRIS data. The untwisting motion is common and often observed in coronal jets \citep{pike98,kamio2010,curdt2012,moore15} and also in jet MHD simulations \citep[e.g.][]{pariat15,wyper18,zhelyazkov18,doyle19}. 

Our active region jet observations show the same picture as coronal hole jets \citep{sterling15} and quiet region \citep{panesar16b} jets. These previous studies also found that such jets are driven by a minifilament eruption, originating from a minifilament-containing flux rope in the base of the jet. The flux rope in these is built up and subsequently triggered to erupt by magnetic flux cancelation at the neutral line. Our example active region jet shows a clear pre-jet minifilament that forms and erupts from the neutral line where continuous flux cancelation is ongoing. The JBP appears on the cancelation neutral line under the erupting minifilament. Furthermore, our active region jet event originates from the  site of clear flux cancelation unlike several other active region jets where flux cancelation is often harder to sort out from ongoing flux emergence \citep[e.g.][]{mulay16,panesar16a,sterling16}. 


Our results support the interpretation that at least some active region coronal jets are miniature analogs of CME-producing eruptions, as both are prepared for eruption and triggered to erupt by magnetic flux cancelation. Coronal jets also carry substantial magnetic, thermal, and kinetic energy, sufficient to locally heat the solar atmosphere.

\section{Conclusion}

Using high-resolution \hri\ images, IRIS slit-jaw images and Mg II rastered spectra, along with AIA data and HMI magnetograms, we conducted a detailed analysis of an isolated active region jet. The active region jet shows a clear minifilament that forms and erupts from the magnetic neutral line where ongoing flux cancelation occurs at the base of the jet. In active region jets, it is often difficult to distinguish between flux emergence and flux cancelation for the cause of the buildup and triggering of a jet, as both processes frequently occur during certain phases of the region’s evolution \citep[e.g.][]{mulay16,sterling16,sterling17}.

Multiple jets originate from the same neutral line, triggered by the ongoing flux cancelation. We infer that flux cancelation gradually builds the shear \citep{balle89} in the magnetic field supporting the minifilament of cool plasma, leading to the formation of a flux rope that contains the minifilament. In accordance with the scenario proposed by  \cite{moore92}, continued flux cancelation eventually triggers the eruption of this flux rope. Further, we conclude that this active region jet is governed by the same key physical processes as  in jets in quiet Sun regions and coronal holes.

Finally, we note that due to the magnetic complexity of active region jets, we cannot rule out that alternative mechanisms may also be at play in some cases. Clear observational cases, such as the one presented in this study, remain relatively rare. To further establish the formation and eruption processes of active region jets, additional such high-resolution observations, particularly as from \hri, are needed.

\begin{acknowledgments}
We thank the referee for the valuable comments that helped improve the paper. We are also grateful to Marc DeRosa  for assistance with the PFSS extrapolations, and to V. Aparna for discussions on potential field extrapolartions. NKP acknowledges support from NASA's SDO/AIA grant (NNG04EA00C) and HSR grant (80NSSC24K0258).  SKT gratefully acknowledges support by NASA contract NNM07AA01C (Hinode). SKT, and NKP
sincerely acknowledge support from HSR grant (80NSSC23K0093) and NSF AAG award (no. 2307505). 
ACS and RLM acknowledge support from NASA's HSR grant. A.Z., M.M. and C.V. thank the Belgian Federal Science Policy Office (BELSPO) for the provision of financial support in the framework of the PRODEX Programme of the European Space Agency (ESA) under contract numbers 4000143743 and 4000134088.
We acknowledge the use of  Solar Orbiter/EUI and  \sdo/AIA/HMI data. AIA is an instrument onboard the Solar Dynamics Observatory, a mission for NASA’s Living With a Star program. Solar Orbiter is a space mission of international
collaboration between ESA and NASA, operated by ESA. The EUI instrument was built by CSL, IAS, MPS, MSSL/UCL, PMOD/WRC, ROB, LCF/IO with funding from the Belgian Federal Science Policy Office (BELSPO/PRODEX PEA 4000134088, 4000112292 and 4000106864); the
Centre National d’Etudes Spatiales (CNES); the UK Space
Agency (UKSA); the Bundesministerium für Wirtschaft und
Energie (BMWi) through the Deutsches Zentrum für Luft- und
Raumfahrt (DLR); and the Swiss Space Office (SSO). This work has made use of NASA ADSABS and Solar Software.
\end{acknowledgments}

%
%



\bibliographystyle{aasjournal}

\end{document}